\documentclass[twocolumn,preprintnumbers,amsmath,amssymb]{revtex4}

\usepackage{lscape}
\usepackage{graphicx}
\usepackage{dcolumn}
\usepackage{bm}
\usepackage{longtable}

\begin{document}
\preprint{APS/}

\title{Intersubband decay of 1-D exciton resonances in carbon nanotubes}

\author{Tobias Hertel$^{1,2}$*, Vasili Perebeinos$^3$ Jared Crochet$^1$,
Katharina Arnold$^4$, Manfred Kappes$^{4,5}$, Phaedon Avouris$^3$}


\affiliation{$^1$Department of Physics and Astronomy, Vanderbilt
University, Nashville, TN 37235, USA} \affiliation{$^2$Vanderbilt
Institute for Nanoscale Science and Engineering, Vanderbilt
University, Nashville, TN 37235, USA} \affiliation{$^3$IBM T.J.
Watson Research Center, Yorktown Heights, NY 10598}
\affiliation{$^4$Institut f\"{u}r Nanotechnologie, Forschungszentrum
Karlsruhe, and Institut f\"{u}r Physikalische Chemie,
Universit\"{a}t Karlsruhe, Karlsruhe, Germany}

\date{\today}

\begin{abstract}
We have studied intersubband decay of $E_{22}$ excitons in
semiconducting carbon nanotubes experimentally and theoretically.
Photoluminescence excitation line widths of semiconducting nanotubes
with chiral indicess $(n,m)$ can be mapped onto a connectivity grid
with curves of constant $(n-m)$ and $(2n+m)$. Moreover, the global
behavior of $E_{22}$ linewidths is best characterized by a strong
increase with energy irrespective of their $(n-m) mod(3)= \pm 1 $
family affiliation. Solution of the Bethe-Salpeter equations shows
that the $E_{22}$ linewidths are dominated by phonon assisted
coupling to higher momentum states of the $E_{11}$ and $E_{12}$
exciton bands. The calculations also suggest that the branching
ratio for decay into exciton bands vs free carrier bands,
respectively is about 10:1.
\end{abstract}

\maketitle

The energetics and dynamics of excited states in carbon nanotubes
(CNTs) receive attention because of their relevance for
non-equilibrium carrier transport in nanotube electronic devices and
their importance for a better understanding of photoluminescence
quantum yields, branching ratios and other fundamental photophysical
properties of CNTs
\cite{tans98,martel98,javey03,freitag06,misewich03,marty06}.

Photoluminescence spectroscopy of semiconducting carbon nanotubes
has been extraordinarily useful not only in determining the
energetics of excitons and their fine structure
\cite{wang05,maultzsch05,mortimer07} but also for establishing
detailed structure-property relationships
\cite{bachilo02,arnold04,shan04,souza05}. Striking family patterns
of tubes characterized by chiral indices $(n,m)$ have been used to
relate first and second subband exciton energies, $E_{11}$ and
$E_{22}$ respectively, and are found to lie on a grid of curves
connecting tubes with constant $(n-m)$ and constant $(2n+m)$
\cite{bachilo02}. Moreover, tubes whose chiral indices are such that
$(n-m) mod(3)=-1$ generally have large $E_{22}/E_{11}$ ratios with
values closer to 2 while those with $(n-m) mod(3)=+1$ have an
$E_{22}/E_{11}$ ratio significantly below 2 \cite{bachilo02,kane03}.
Similar patterns, such as for strain and temperature dependence of
exciton transitions for example \cite{arnold04,shan04,souza05}, also
indicate a strong sensitivity to chirality and tube structure.
However, evidence for family relationships in the dynamical
properties of excited states is scarce, and how strongly electronic
or vibronic processes contribute to the exciton dynamics is not yet
resolved.

Here, we examine family connectivities of exciton interband decay
rates and their implications for exciton dynamics in semiconducting
SWNTs. The observed family patterns in combination with solutions of
the Bethe-Salpeter equation \cite{perebeinos04,perebeinos05b} allow
to identify zone boundary phonon scattering as the dominant
facilitator for interband relaxation. Our calculations moreover show
that the branching ratio for decay of the $E_{22}$ exciton into the
first subband \textit{e-h} pair continuum and the lower lying
exciton bands is about 1:10.

Single-wall carbon nanotube suspensions are prepared from commercial
nanotube material (HiPCO, CNI Houston). We used two types of
samples: type (A) made from soot ultrasonically dispersed and
ultracentrifuged in $\rm{D_2O}$ using SDS as surfactant
\cite{oconnel02} which is here referred to as SDS-SWNT and type (B)
made from nanotubes of the same raw material after isopycnic
fractionation from $\rm{H_2O}$ nanotube suspensions with
sodium-cholate as surfactant \cite{arnold05,crochet07} here referred
to as SC-SWNT.

\begin{figure}
\includegraphics[width=8cm]{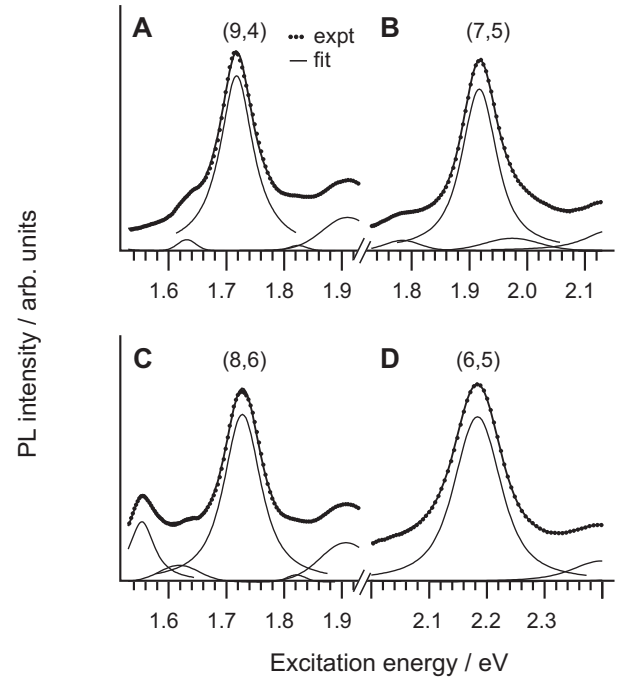}
\caption{Four selected photoluminescence excitation spectra for
emission from an (A) type sample at a) 1109\,nm, b) 1031\,nm, c)
1180\,nm and d) 983\,nm. The center peaks in these spectra
correspond to the $E_{22}$ exciton resonances of (9,4), (7,5), (8,6)
and (6,5) nanotubes respectively. The thin solid line is the fit
used to determine linewidths.} \label{fig1}
\end{figure}

Photoluminescence excitation spectra from SDS-SWNTs were measured in
the range of 800\,nm - 1700\,nm with a Bruker Equinox 55S/FRA106 FT
Raman spectrometer equipped with a liquid nitrogen cooled germanium
detector \cite{lebedkin03}. SC-SWNTs were studied with  a Jobin
Yvon/Horiba Fluorolog-3 FL3-111 spectrophotofluorometer equipped
with a liquid nitrogen cooled InGaAs detector. PL spectra were
recorded with 8 nm excitation spectral slits and a
$20\,\rm{cm}^{-1}$ resolution of the FT-Raman spectrometer and with
10 nm exitation slit width in the case of measurements performed
with the Flurolog. Both sample types are found to have very similar
excitation line-widths. Small systematic deviations are mostly
attributed to the different resolution of spectrometers used for
both experiments.

A set of selected photoluminescence excitation spectra is reproduced
in Fig. \ref{fig1} where we plot PL intensities at 1109\,nm,
1031\,nm, 1180\,nm and 983\,nm as a function of excitation
wavelength. The center peaks in these spectra correspond to the
$E_{22}$ exciton resonances of (9,4), (7,5), (8,6) and (6,5)
nanotubes respectively. To extract the width of $E_{22}$ exciton
resonances we used a multi-Voigt peak, non-linear least squares fit
routine with polynomial background subtraction. The routine is part
of the IGOR software package (Wavemetrics) and uses the
Levenberg-Marquardt algorithm for minimization of chi-square. The
results of such a fit to select datasets are likewise shown in Fig.
\ref{fig1} as solid lines. The estimated error margins are
significantly larger than the confidence bands of the nonlinear
least squares fit and are based on the reproducibility of the fit if
results from different datasets are compared including mismatch of
the results obtained for SDS- and SC-SWNT samples. For the following
discussion we corrected data for instrumental resolution. The
linewidths used later on, thus include inhomogeneous broadening
which is estimated to be on the order of $10$\,meV based on a
comparison of linewidths for $E_{11}$ emission features from vacuum-
and surfactant suspended SWNTs (see for example ref.
\cite{inoue06}).

\begin{figure}
\includegraphics[width=8cm]{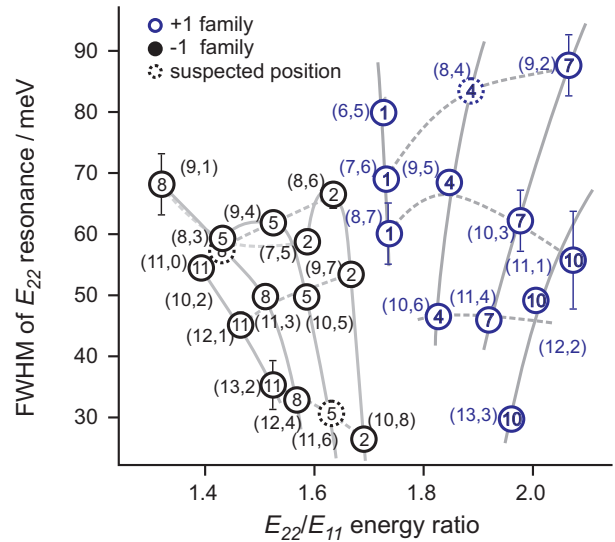}
\caption{Photoluminescence excitation linewidths versus the
$E_{22}/E_{11}$ ratio. The $(n-m)$ family affiliation is indicated
by the corresponding numbers within the datapoints. Horizontal lines
connect tubes with the same $(2n+m)$.} \label{fig2}
\end{figure}

To identify the mechanisms leading to the observed linewidths we
begin with an investigation of their chirality and energy
dependence. If we plot $E_{22}$ linewidths versus their respective
$h\nu_{22}/h\nu_{11}$ energy ratios, for example, the data follows a
striking family pattern where individual data points can be
connected by a grid of intersecting curves with constant $(n-m)$ and
$(2n+m)$ (see Fig. \ref{fig2}. The numbers inside the datapoints
give the  $(n-m)$ family affiliation. Family connectivities with
patterns similar to this one have previously been found for the
photoluminescence excitation maps of semiconducting SWNTs
\cite{bachilo02} but also for the temperature and strain dependence
of exciton line positions \cite{arnold04,shan04,souza05}. The lines
with constant $(n-m)$ converge towards 1.7, the large-tube limit of
the energy ratio $h\nu_{22}/h \nu_{11}$. The overall trend is for
smaller tubes to have larger linewidths, a trend which is reversed
only by the smallest tubes of the -1 family studied here,
specifically the (8,3) and (7,5) tubes.

\begin{figure}
\includegraphics[width=8cm]{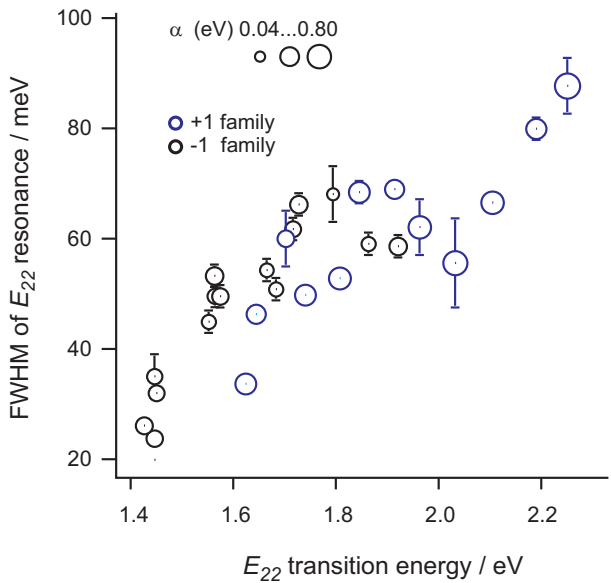}
\caption{The global behavior of $E_{22}$ linewidths is best
characterized by a pronounced energy dependence with no discernible
influence of $+1$ or $-1$ family affiliation.} \label{fig3}
\end{figure}

In order to identify the microscopic process responsible for the
observed family connectivity and for the width of $E_{22}$ exciton
resonances we now search for correlations of linewidths with
established structural and electronic properties such as chirality,
tube diameter, $E_{11}$ or $E_{22}$ transition energies, the
$(h\nu_{22}-h\nu_{11})$ or $(h\nu_{22}-\Delta_{11}^\Gamma)$ energy
differences, where $\Delta_{11}^\Gamma$ refers to the free $e-h$
pair generation threshold. Aside from the family connectivity
discussed above we find that the by far clearest and strongest
correlation of $E_{22}$ resonance widths is with the energy of the
$E_{22}$ resonance itself (see Fig. \ref{fig3}). We also note that
we find no evidence for an explicit dependence of linewidths on the
$\pm 1$ family affiliation, and the larger linewidths of the +1
family tubes in Fig. \ref{fig2} can be exclusively attributed to a
correlation with their higher $E_{22}$ energies.

\begin{figure}
\includegraphics[width=8cm]{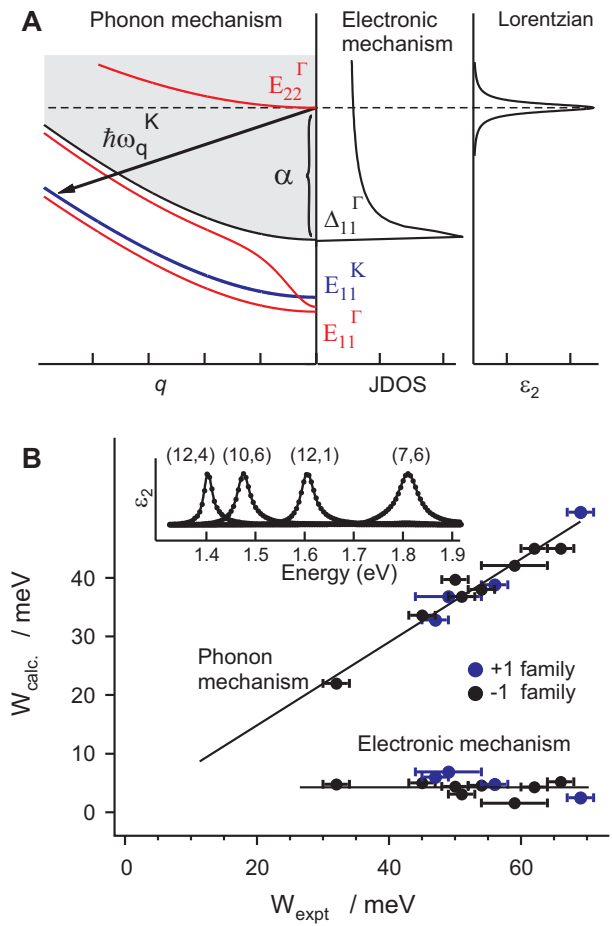}
\caption{A) Schematic illustration of two possible mechanisms
leading to decay of the $E_{22}$ exciton: a) phonon mediated
coupling to the $E_{11}$ exciton bands (left side) and b) decay by
resonant coupling to the free $e-h$ continuum (middle). B)
comparison of $W_{expt}$ experimental $E_{22}$ linewidths with
calculated ones ($W_{calc.}$). Only the phonon-mechanism shows a
clear correlation between experiment and theory. The inset shows a
selection of calculated excitation spectra with clearly increasing
linwidths at higher excitation energies.} \label{fig4}
\end{figure}

To explore the pronounced energy dependence of resonance linewidths
in further detail we consider the two microscopic mechanisms that
can lead to energy relaxation from the $E_{22}$ state. One is
referred to as phonon-mechanism because it describes phonon mediated
coupling of $E_{22}^\Gamma$, the $\Gamma$-point exciton to lower
lying exciton states of the $E_{11}$ and $E_{12}$ manifolds at
energies $(h\nu_{22}-\hbar\omega_q)$ with momentum $q$ (see left
panel of Fig. \ref{fig4}A). The second possible mechanism describes
direct coupling of the $E_{22}$ state to the $\Gamma$-point
continuum of free $e-h$ pair excitations with its threshold at
$\Delta_{11}^\Gamma$. If the rate of decay $W_{el}$ for the
electronic process is approximated using Fermi's Golden Rule we
obtain

\begin{equation}
W_{el}\propto \left| \langle
\psi_{22}|V_{el}|\psi_{e-h}\rangle\right|^2 \rho_{e-h}(h\nu_{22})
\label{eq1}
\end{equation}

\noindent where $\psi_{22}$ and $\psi_{e-h}$ are wavefunctions of
the second subband exciton and the continuum state in the first
subband into which the $E_{22}$ exciton decays. $V_{el}$ is the
coupling term and $\rho_{e-h}$ is the joint density of states of the
final state continuum. Accordingly, one would expect the dependence
of $W_{el}$ on the JDOS at the energy $h\nu_{22}$ to lead to a
pronounced dependence of the FWHM on the band alignment of the
$E_{22}$ state with respect to the free carrier band edge. By a
comparison of linewidths with the alignment parameter
$\alpha=(h\nu_{22}^\Gamma-\Delta_{11}^\Gamma)$ (see Fig.
\ref{fig4}a) we should therefore be able to assess the importance of
the electronic process for $E_{22}$ energy relaxation.

The alignment parameter $\alpha$ is determined using a combination
of experimental data and theoretical considerations. Recent
calculations suggest that exciton binding energies $E_{b,11}$ scale
linearly with the renormalized free-carrier gap \cite{capaz06}. A
similar scaling consequently also characterizes the relationship of
$E_{11}$ binding energies with the corresponding optical transition
energies. By comparison of calculated \cite{capaz06} with
experimental $E_{11}$ binding- and transition energies $E_{b,11}$
and $h\nu_{11}$, respectively \cite{wang05, maultzsch05}, we find
$E_{b,11}=0.31\,h\nu_{11}$ and the free carrier gap becomes
$\Delta_{11}^\Gamma=1.31\,h\nu_{11}$. We can thus determine the
alignment parameter $\alpha(n,m)$. From this we find that the
$E_{22}$ state is in resonance with the free carrier continuum for
all tubes studied here. Moreover, the absence of any correlation of
linewidths with $\alpha$ (see Fig. \ref{fig3}) suggests that
coupling to the free carrier continuum is weak.

To put these findings on a more quantitative basis we solved the
Bethe-Salpeter equation, as described in Ref. \cite{perebeinos04},
using a $\pi$-orbital tight-binding basis set with a hopping
parameter $t_0=3.0$ eV. This allows to determine linewidths from the
frequency dependence of the absorption coefficient $\epsilon_2$
\cite{perebeinos04}. In the electronic decay mechanism, the $E_{22}$
exciton couples to the free $e-h$ continuum, as schematically shown
in the right panel of Fig.~\ref{fig4}A, leading to a Lorentzian
absorption spectrum \cite{spataru04,perebeinos04}. Several states of
the Bethe-Salpeter equation (BSE) contribute to the $E_{22}$ exciton
absorption. We choose a unit cell of about 500 nm length, large
enough to resolve the spectral width of the $E_{22}$ peak. The
linewidths resulting from the fit of Lorentz profiles to the
calculated absorption spectra are shown in Fig.~\ref{fig4}B.
However, the predicted broadening due to the electronic mechanism
is, on average, a factor of 10 smaller than the experimental width
and no correlation of calculated and experimental values is found.

The second decay mechanism discussed here involves exciton-phonon
coupling, which allows the $E_{22}$ exciton to decay to a lower
energy state by phonon emission. Unlike in the electronic mechanism,
which requires the initial exciton state and final electron-hole
pair states to have the same angular and longitudinal momenta,
namely $q=0$,  in the exciton-phonon mechanism, the phonon can have
both finite angular and longitudinal momenta resulting in a final
state with finite momentum. To explore this phonon assisted $E_{22}$
decay mechanism, we used the Su-Schrieffer-Heeger model \cite{Su}
for the electron-phonon interaction, with a matrix element
$t=t_0-g\delta R_{C-C}$ dependent on the change of the nearest
neighbor C-C distance $\delta R_{C-C}$. We use $g=5.3$ eV/\AA.
Following Ref. \cite{perebeinos05b}, we calculate the exciton-phonon
coupling using exciton wavefunctions from the BSE solution. We then
use the Golden rule to calculate the lifetime of each BSE state,
which contributes to the $E_{22}$ absorption spectra including the
exciton-phonon coupling \cite{perebeinos05b}. The $E_{22}$ spectrum
is then again fit with a Lorentzian profile with the resulting
$E_{22}$ FWHM plotted in Fig.~\ref{fig4}B to illustrate the clear
correlation of theory and experiment. In our calculations, we
allowed for $E_{22}$ exciton coupling with all phonons and found
that the $E_{22}$ exciton primarily decays into the
doubly-degenerate dark $E_{11}$ exciton \cite{perebeinos05} with a
finite angular momentum via the zone-boundary $K$-optical phonon.
The slope of the straight line fit to the calculated linewidths
using the phonon-mechanism in Fig.~\ref{fig4}B is about 30\% smaller
than experimental data. Some of this difference may be due to
phonon-assisted decay into sigma states \cite{louie94} as well as to
thermal broadening by acoustic phonons on the order of a few meV,
which are not included in our theory.

The absence of any clear correlation of linewidths and the family
affiliation in the experiment is also consistent with the phonon
decay mechanism. We note, that there is a difference in the
electron-phonon coupling for the two families in the single particle
model. \cite{jiang07} However, the exciton binding energies also
exhibit a family effect related to their effective masses
\cite{perebeinos04}. As a result, the stronger bound excitons have
larger exciton-phonon coupling \cite{perebeinos05b}, which in effect
compensates for the difference in the electron-phonon coupling in
the single particle picture \cite{jiang07}. However, the complex
family correlations seen experimentally in Fig.~\ref{fig2} cannot be
understood at this level of theory and may provide motivation to
study these effects using more sophisticated approaches, for
example, including the $\sigma-\pi$ orbital hybridization.

In conclusion, we find that linewidths of the second subband
$E_{22}$ excitons in semiconducting carbon nanotubes exhibit
pronounced family connectivities and strongly depend on the $E_{22}$
transition energy. Our calculations show that the dominant
scattering process to the lower lying dark $E_{11}$ exciton subband
is facilitated by coupling via zone boundary phonons from the
$E_{22}$ resonance. The linewidths are found to be consistent with
an $e-ph$ coupling strength of about 5\,eV/\AA. In contrast, the
coupling to the free carrier continuum is comparatively small and
leads to a branching ratio for decay into the lower exciton versus
free carrier bands of roughly 10:1. Future studies will need to
clarify finer effects of the family connectivity of the relaxation
rates.

\end{document}